\documentclass[reprint,twocolumn,pre,showpacs,amsmath,amssymb,aps]{revtex4-1}

\usepackage{amsmath}
\usepackage{natbib}	
\usepackage{graphicx,color,rotating}
\usepackage[latin1]{inputenc}
\usepackage{textcomp}
\usepackage{dcolumn}% Align table columns on decimal point
\usepackage{bm}     % bold math
\usepackage{upgreek}
\usepackage{subdepth}  			% Added by JTK (aligns sub-indices)
\usepackage{multirow}
\usepackage[latin1]{inputenc}
\usepackage{hyperref}						% Package for hyperreferences
\usepackage{xcolor}						% Allows for coloring of text
\hypersetup{							% Setup of the Hyperref package
    colorlinks,
    linkcolor={blue!80!black},
    citecolor={blue!80!black},
    urlcolor={blue!80!black}
}
\newcommand{\mr}[1]{\ensuremath{\mathrm{#1}}}
\renewcommand{\vec}[1]{\bm{#1}}
\newcommand{\ee}{\mathrm{e}}

\newcommand{\ii}{\mathrm{i}}
\newcommand{\dm}{\mathrm{d}}
\newcommand{\avr}[1]{\big\langle #1 \big\rangle}

\newcommand{\pbar}{\bar{p}}
\newcommand{\Pbar}{\bar{P}}
\newcommand{\pp}{\partial}

\newcommand{\psiact}{\psi_\mr{act}}

\newcommand{\psiactbox}{\psiact^\mr{box}}
\newcommand{\psiactnum}{\psiact^\mr{num}}

\newcommand{\chidelta}{\chi_\delta}

\newcommand{\chibox}{\chi_\mr{box}}

\newcommand{\nablabf}{\boldsymbol{\nabla}}

\newcommand{\ie}{\textit{i.e.}}

\newcommand{\etal}{\textit{et~al.}}

\newcommand{\qmarks}[1]{``{#1}"}

\newcommand*{\plimsoll}{{\ensuremath{-\kern-4pt{\ominus}\kern-4pt-}}}

\newcommand{\cfl}{c_\mr{fl}}

\newcommand{\dact}{d_\mr{act}}
\newcommand{\dactti}{\tilde{d}_\mr{act}}

\newcommand{\dd}{\mr{d}}

\newcommand{\eee}{\vec{e}}

\newcommand{\FFFrad}{\vec{F}_\mathrm{rad}}
\newcommand{\FFFradbar}{\bar{\vec{F}}_\mathrm{rad}}

\newcommand{\kc}{k_\mathrm{c}}
\newcommand{\kcbar}{\bar{k}_\mathrm{c}}

\newcommand{\kx}{k_x}

\newcommand{\deltas}{\delta_\mr{s}}

\newcommand{\ks}{k_\mathrm{s}}

\newcommand{\nnn}{\vec{n}}

\newcommand{\phat}{\hat{p}}

\newcommand{\rrr}{\vec{r}}

\newcommand{\SSSac}{\vec{S}_\mathrm{ac}}
\newcommand{\SSSacbar}{\bar{\vec{S}}_\mathrm{ac}}

\newcommand{\uuu}{\vec{u}}

\newcommand{\calD}{\mathcal{D}}
\newcommand{\calE}{\mathcal{E}}

\newcommand{\calF}{\mathcal{F}}

\newcommand{\calGbar}{\bar{\mathcal{G}}}

\newcommand{\cO}{c_0}

\newcommand{\Ekin}{E_\mathrm{kin}}
\newcommand{\Ekinbar}{\bar{E}_\mathrm{kin}}

\newcommand{\Epot}{E_\mathrm{pot}}
\newcommand{\Epotbar}{\bar{E}_\mathrm{pot}}

\newcommand{\Eac}{E_\mathrm{ac}}

\newcommand{\Eacbar}{\bar{E}_\mathrm{ac}}

\newcommand{\fbar}{\bar{f}}

\newcommand{\kbar}{\bar{k}}

\newcommand{\kapfl}{\kappa_\mathrm{fl}}

\newcommand{\Lc}{L_\mathrm{cr}}

\newcommand{\Lx}{L_x}
\newcommand{\Ly}{L_y}
\newcommand{\Lz}{L_z}

\newcommand{\Lact}{L_\mathrm{act}}

\newcommand{\Lactti}{\tilde{L}_\mathrm{act}}

\newcommand{\Rbar}{\bar{R}}

\newcommand{\Ubar}{\bar{U}}

\newcommand{\Urad}{U_{\mathrm{rad}_{}}}
\newcommand{\Uradbar}{\bar{U}_{\mathrm{rad}_{}}}

\newcommand{\etafl}{\eta_\mr{fl}}

\newcommand{\etaBfl}{\eta^\mathrm{b}_\mathrm{fl}}

\newcommand{\Gammabar}{\bar{\Gamma}}

\newcommand{\Gamfl}{\Gamma_\mathrm{fl}}
\newcommand{\Gamblbar}{\bar{\Gamma}_\mr{bl}}

\newcommand{\rhofl}{\rho_\mathrm{fl}}

\newcommand{\xti}{{\tilde{x}{}}}

\newcommand{\omegabar}{\bar{\omega}}
\newcommand{\Omegabar}{\bar{\Omega}}
\newcommand{\dOmegabar}{\partial\bar{\Omega}}

\newcommand{\kO}{k_{0}}

\newcommand{\SIcm}{\textrm{cm}}
\newcommand{\SIum}{\upmu\textrm{m}}

\newcommand{\SImm}{\textrm{mm}}
\newcommand{\SImum}{\textrm{\textmu{}m}}

\newcommand{\SIpN}{\textrm{pN}}

\newcommand{\nn}{\nonumber}
\newcommand{\beq}[1]{\begin{equation} \eqlab{#1}}
\newcommand{\eeq}{\end{equation}}
\newcommand{\bsub}{\begin{subequations}}
\newcommand{\esub}{\end{subequations}}
\def\bal#1\eal{\begin{align}#1\end{align}}
\def\balat#1#2\ealat{\begin{alignat}{#1} #2 \end{alignat}}
\def\bsublab#1#2\esublab{\bsub \eqlab{#1} #2 \esub}
\def\bsubal#1#2\esubal{\bsublab{#1}\begin{align}#2\end{align} \esublab}% begin/end align with a,b,c-equation labels
\def\bsubalat#1#2#3\esubalat{\bsublab{#1} \begin{alignat}{#2} #3 \end{alignat} \esublab}

\newcommand{\eqlab}[1]{\label{eq:#1}}
\renewcommand{\eqref}[1]{Eq.~(\ref{eq:#1})}
\newcommand{\eqnoref}[1]{(\ref{eq:#1})}

\newcommand{\eqsref}[2]{Eqs.~(\ref{eq:#1}) and~(\ref{eq:#2})}

\newcommand{\figref}[1]{Fig.~\ref{fig:#1}}

\newcommand{\figsref}[2]{Figs.~\ref{fig:#1} and~\ref{fig:#2}}
\newcommand{\figlab}[1]{\label{fig:#1}}
\newcommand{\appref}[1]{Appendix~\ref{sec:#1}}

\newcommand{\secref}[1]{Section~\ref{sec:#1}}

\newcommand{\seclab}[1]{\label{sec:#1}}
\newcommand{\tabref}[1]{Table~\ref{tab:#1}}

\newcommand{\tablab}[1]{\label{tab:#1}}

\newcommand{\ord}[1]{\mathcal{O}({#1})}
\newcommand{\ex}{\eee_{x}}

\newcommand{\grad}{\boldsymbol{\nabla}}

\newcommand{\lap}{\nabla^2}

\newcommand{\intinf}{\int_{-\infty}^\infty}

\renewcommand{\Re}{\mathrm{Re}}
\renewcommand{\Im}{\mathrm{Im}}

\begin{document}
%\preprint{Preprint identifier}

\title{Theory of acoustic trapping of microparticles in capillary tubes}

\author{Jacob S. Bach}
\email{jasoba@fysik.dtu.dk}
\affiliation{Department of Physics, Technical University of Denmark,\\ DTU Physics Building 309, DK-2800 Kongens Lyngby, Denmark}

\author{Henrik Bruus}
\email{bruus@fysik.dtu.dk}
\affiliation{Department of Physics, Technical University of Denmark,\\
DTU Physics Building 309, DK-2800 Kongens Lyngby, Denmark}

\date{29 November 2019}

\begin{abstract}

We present a semi-analytical theory for the acoustic fields and particle-trapping forces in a viscous fluid inside a capillary tube with arbitrary cross section and ultrasound actuation at the walls. We find that the acoustic fields vary axially on a length scale proportional to the square root of the quality factor of the two-dimensional (2D) cross-section resonance mode. This axial variation is determined analytically based on the numerical solution to the eigenvalue problem in the 2D cross section. The analysis is developed in two steps: First, we generalize a recently published expression for the 2D standing-wave resonance modes in a rectangular cross section to arbitrary shapes, including the viscous boundary layer. Second, based on these 2D modes, we derive analytical expressions in three dimensions for the acoustic pressure, the acoustic radiation and trapping force, as well as the acoustic energy flux density.  We validate the theory by comparison to three-dimensional numerical simulations.

\end{abstract}

% \pacs{43.25.Qp, 43.20.Fn, 43.20.+g, 47.35.Rs}

% 43 Acoustics
%   43.20.+g: General linear acoustics
%   43.20.Fn: Scattering of acoustic waves
%   43.20.Ks: Standing waves, resonance, normal modes
%	47.35.Rs: Sound waves in fluids
%   43.25.Gf: Standing waves; resonance
%   43.25.-x: Nonlinear acoustics
%	43.25.Nm: Acoustic streaming
%   43.25.Qp: Radiation pressure
%   43.35.Ty: Other physical effects of sound
%   43.20.Bi: Mathematical theory of wave propagation
%	43.80.-n, 43.80.+p: Ultrasound application to biology
% 47 Fluid dynamics
%   47.15.-x: laminar
%	47.35.Rs: Sound waves in fluids

%\keywords{Suggested keywords} Use showkeys class to display keywords

\maketitle

% Main text

\section{Introduction}
\seclab{intro}

Acoustophoresis is the acoustically induced migration of particles. During the past few decades the scientific field of microscale acoustofluidics has emerged, where this phenomenon is exploited for controlled handling of microparticles. Microscale acoustophoresis is gentle, label-free, and contact-less, and therefore useful for bioanalytics in lab-on-a-chip technologies. Examples include particle separation \cite{Laurell2007, Tan2009, Lenshof2010, Ding2014}, concentration of red blood cells \cite{Lenshof2009}, iso-acoustic focusing of cells \cite{Augustsson2016}, acoustic tweezing \cite{Shi2009a, Riaud2017, Gong2019}, and cell patterning \cite{Collins2015, Collins2018}. One particularly prominent acoustofluidic application is acoustic trapping of suspended microparticles against an external flow in cheep, disposable glass capillary tubes \cite{Lilliehorn2005, Evander2007, Hammarstrom2010, Lei2013, Mishra2014, Gralinski2014}, which has been used for fast biological assays \cite{Grundy1993, Tenje2014} and for trapping of sub-micrometer particles by use of larger trapped seed particles \cite{Hammarstrom2012, Evander2015}. In these systems, a piezoelectric transducer is attached to the capillary tube and driven at MHz frequencies to generate a standing-wave resonance mode localized inside the capillary tube above the transducer.

The physics behind acoustophoresis is primarily described by two time-averaged forces acting on the suspended particles. First, due to differences in density and compressibility between the particles and the carrier fluid, the particles experience the acoustic radiation force, which scales with the particle volume \cite{King1934, Yosioka1955, Gorkov1962, Doinikov1997, Settnes2012, Karlsen2015} and tends to focus particles. Second, due to time-averaged momentum fluxes induced by the acoustic fields, a steady acoustic streaming flow is generated, and suspended particles therefore experience a drag force, which scales with the particle radius and tends to mix particles \cite{LordRayleigh1884, Eckart1948, Bach2018}.

Both the acoustic radiation force and the acoustic streaming are important in the acoustic trap. Aside from the one-dimensional (1D) or two-dimensional (2D) focusing in the cross section due to the transverse acoustic radiation force, the axial variations in the acoustic fields also give rise to an axial acoustic radiation force, or trapping force. Furthermore, acoustic streaming in the plane parallel to the transducer surface is often observed above the edges of the transducer, strongly affecting the trapping characteristics \cite{Hammarstrom2012, Lei2013}. The acoustic trapping in capillaries is therefore a three-dimensional (3D) problem, see \figref{3D_fig}, which complicates both the experimental characterization and the theoretical analysis needed for further development.

 \begin{figure}[t]
 \includegraphics[width=1\columnwidth]{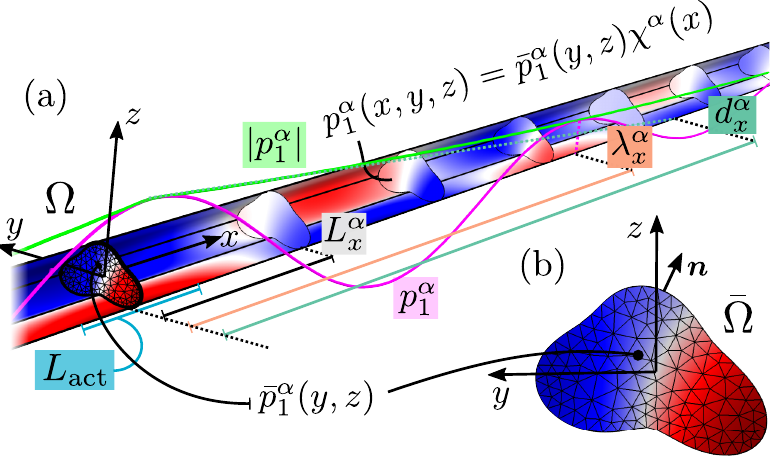}
 \caption{\figlab{3D_fig} A computed pressure resonance mode (red/blue for high/low pressure) in a capillary tube of arbitrary cross section. (a) The complex-valued pressure mode $p_1^\alpha(x,y,z)$ in the 3D tube $\Omega$ is the product of the 2D pressure mode $\pbar_1^\alpha(y,z)$ in the cross section $\Omegabar$ and the axial dependency $\chi^\alpha(x)$. It is excited by an actuation confined to a region of length $\Lact$. The magenta and green curves represent the real part of $p_1^\alpha$ and the magnitude $|p_1^\alpha|$, respectively. The relevant length scales in the $x$ direction are shown: the wave length $\lambda_x^\alpha$, the decay length $d_x^\alpha$, and the characteristic length scale $\Lx^\alpha$, see \eqsref{Lx}{dx_lambdax}. (b) The pressure $\pbar_1^\alpha(y,z)$ in the 2D cross section $\Omegabar$ with the surface normal vector $\nnn$ and the mesh used in the numerical simulations.}
 \end{figure}

In this work, we present a method to semi-analytically calculate the 3D acoustic pressure in a capillary tube of arbitrary cross section actuated in an axially confined region of length $\Lact$ at the walls, see \figref{3D_fig}. Based on the either analytical or numerical solution to the 2D eigenvalue problem in the cross section, we derive analytical expressions in three dimensions for the acoustic pressure, the acoustic radiation force, the acoustic energy flux density, and the ratio between the axial and transverse acoustic trapping force. In particular, we show that for a 2D resonance mode $\alpha$, the axial component of the radiation force is proportional to $\sqrt{\Gammabar^\alpha}$, where $\Gammabar^\alpha$ is the damping coefficient of the resonance mode $\alpha$. In the special case of a 1D standing pressure wave in the cross section, our results agree with Woodside \etal~\cite{Woodside1997}, who obtained an analytical expression for the axial radiation force being proportional to the axial gradient of the acoustic energy density $\Eac$. However, whereas they left $\Eac$ undetermined, we calculate it analytically.

We validate our analytical results by direct 3D numerical simulations. Recent contributions in the 3D numerical modeling of capillary tubes include Gralinski~\etal~\cite{Gralinski2014} who modeled a circular capillary tube with fluid and glass, Lei \etal~\cite{Lei2013}, who modelled the fluid domain of a capillary tube and found four in-plane streaming rolls, and Ley and Bruus~\cite{Ley2017}, who took into account absorption of outgoing waves in both the glass and the fluid. Also the piezoelectric transducer may be included in a full-device simulation as done by Skov \etal~\cite{Skov2019}. In the numerical validation of this paper, we model the fluid domain with a prescribed movement of the fluid-solid interface and implement a perfectly mached layer (PML) to absorb outgoing waves.

The paper is organized as follows: We present the governing equations in \secref{Goveq}, and in \secref{numeric_impl} we describe the numerical implementation used for validation of the presented theory. In \secref{theory}, we generalize our previous analytical results for the acoustic pressure in rectangular cross sections~\cite{Bach2019} to arbitrary cross sections. Using the residue theorem, we derive the axial dependency of the  3D acoustic pressure. We proceed in \secref{envelope_calc} by calculating the axial dependency of the pressure in the case of a box-shaped actuation, and in \secref{numerics3D} we validate the analytical results by 3D numerical simulations. In \secref{physical}, we present analytical expressions for physical time-averaged quantities such as the axial radiation force and the axial energy flux density. Finally, we discuss our results in \secref{discussion} and conclude in \secref{conclusion}.

\section{Governing equations}
\seclab{Goveq}

The physical displacement $\uuu^0_\mr{phys}(\rrr,t)$ of the fluid-solid interface oscillates harmonically with the angular frequency $\omega=2\pi f$ and induces the physical pressure field $p_\mr{phys}(\rrr,t)$ in the fluid. These fields are represented as the real part of the complex-valued linear perturbations $\uuu^0_1$ and $p_1$,
\bsubal{perturbation}
\uuu^0_\mr{phys}(\rrr,t) &= \Re\Big[\uuu_1^0(\rrr) \ee^{-\ii\omega t}\Big],\\
p_\mr{phys}(\rrr,t) &= \Re\Big[p_1(\rrr) \ee^{-\ii\omega t}\Big].
\esubal
In a fluid of dynamic viscosity $\etafl$, bulk viscosity $\etaBfl$, isentropic compressibility $\kapfl$, and mass density $\rhofl$, 
%ZZZ are 
the acoustic fields are characterized by the compressional wave number $\kc$ with real part $k_0=\frac{\omega}{\cfl}$, the bulk damping coefficient $\Gamfl$, the shear wave number $\ks$, and the viscous boundary-layer width $\deltas$ \cite{Karlsen2015, Bach2018, Skov2019},
 \bsubal{kc_ks}
 \eqlab{kc_Gamma}
 \kc &=\Big(1+\ii\frac12 \Gamfl \Big)k_0, &
 \Gamfl &=\bigg(\frac43+\frac{\etaBfl}{\etafl}\bigg)\etafl\kapfl \omega,\\
 \ks &=\frac{1+\ii}{\deltas}, &
 \deltas &=\sqrt{\frac{2\etafl}{\rhofl \omega}},
 \esubal
where $\ii=\sqrt{-1}$ is the imaginary unit. In this work, we assume that the viscous boundary layer is much thinner than the acoustic wave length, as is the case in most acoustofluidic applications,
 \beq{kdelta}
 k_0\deltas\ll 1, \qquad \Gamfl=\frac12 \bigg(\frac43+\frac{\etaBfl}{\etafl}\bigg)(k_0\deltas)^2 \ll 1.
 \eeq
The acoustic pressure $p_1$ satisfies the Helmholtz equation with the compressional wave number $\kc$ and with the boundary-layer boundary condition recently derived in Ref.~\cite{Bach2018}, valid for walls having a curvature radius much larger than the viscous boundary layer width $\deltas$,
 \bsubal{p1_gov}
 \eqlab{p1_gov_helmholtz}
 &\lap p_1+\kc^2 p_1=0,
 & &\rrr\in\Omega,
 \\
 \eqlab{p1_gov_bc}
 & \calD_\perp p_1=k_0^2 U_{1\perp}(\rrr),
 & & \rrr\in\pp\Omega,
 \\ \eqlab{p1_gov_D_def}
 &\calD_\perp =\pp_\perp +\frac{\ii}{\ks}\big(\kc^2 +\pp_\perp^2\big), & &\rrr\in\partial\Omega,
 \\
 \eqlab{p1_gov_U_def}
 &U_{1\perp}(\rrr)= \frac{\rhofl \cfl^2}{1-\ii\Gamfl}\Big(-\nnn - \frac{\ii}{\ks}\nablabf\Big)\cdot\uuu_1^0,
 & &\rrr\in\partial\Omega.
 \esubal
Here, the subscript $\perp$ represents the \emph{inward} direction $(-\nnn)$ opposite to the outward-pointing normal vector $\nnn$, and $U_{1\perp}(\rrr)$ is the effective actuation function defined in terms of the physical interface displacement $\uuu_1^0$ of the fluid-solid interface $\partial\Omega$. Finally, we write the following standard time-averaged acoustic quantities, all defined in terms of the pressure $p_1$: The acoustic potential energy density $\Epot$, the acoustic kinetic energy density $\Ekin$, the acoustic mechanical energy density $\Eac$, the acoustic radiation potential $\Urad$ for a suspended spherical particle of radius $a$, the acoustic radiation force $\FFFrad$, and the acoustic energy flux density $\SSSac$,
 \bsubal{Energies_general}
 \eqlab{Epot_Ekin}
 \Epot&=\frac14 \kapfl |p_1|^2,&   \Ekin&=\frac14 \kapfl k_0^{-2} |\grad p_1|^2, \\
 \eqlab{Eac_Sac}
 \Eac&=\Epot+\Ekin,&   \SSSac &=\frac{1}{2\rhofl\omega}\Im\big(p_1^* \grad p_1\big), \\
 \eqlab{Frad_Urad}
 \FFFrad&= -\grad \Urad, & \Urad&=\frac43\pi a^3\Big[f_0 \Epot-\frac{3}{2}\:f_1 \Ekin\Big].
 \esubal
Here, $f_0$ and $f_1$ are the monopole and dipole scattering coefficients that are real-valued because we consider particles with radius $a$ much larger than both the viscous and the thermal boundary-layer thickness \cite{King1934, Doinikov1997, Settnes2012, Karlsen2015}. Furthermore,  in \eqref{Eac_Sac} \qmarks{Im} and \qmarks{$*$} denotes imaginary part and complex conjugation, respectively.

\section{Numerical validation of the theoretical results}
\seclab{numeric_impl}

We validate numerically our theoretical results for the key theoretical field, the pressure $p_1$, using the weak form PDE module in COMSOL Multiphysics~\cite{Comsol54} as described in  Refs.~\cite{Muller2012,  Bach2018, Skov2019}, see also an example COMSOL script given in the supplemental material of Ref.~\cite{Muller2015}. This validation is carried out in both a 3D and a 2D version solving the harmonically-driven problem~\eqnoref{p1_gov} using the COMSOL \qmarks{\texttt{Stationary study}}. Moreover, as explained in \secref{theory}, a main result of this work is that we can express the 3D pressure in terms of 2D pressure eigenmodes and eigenvalues, which we compute numerically using the COMSOL \qmarks{\texttt{Eigenvalue study}}. In the numerical simulations we use Lagrangian shape functions of quartic order, and the parameters listed in \tabref{params}.

For the numeric validation, we choose a capillary with the generic cross section shown in \figref{3D_fig}. This cross section has a linear size of around $2\Lc$, and its boundary $\pp\Omegabar$ is given by the wavy parametric curve $[y(s),z(s)]$ with $s\in [0;2\pi]$, defined by $y(s)=\Lc h_\mathrm{cr}(s)\cos(s)$ and $z(s)=0.9\Lc h_\mathrm{cr}(s)\sin(s+0.2)$, where $h_\mathrm{cr}(s)=1+0.15\sin(2s+1.5)-0.2\sin(3s)$.

The mesh is chosen to resolve the pressure on the relevant length scales. It is created as a 2D triangular mesh in the cross section with mesh size $\frac13\, \Lc$ in the bulk and $\frac16 \Lc$ at the boundary, see \figref{3D_fig}(b). The 3D mesh is generated by sweeping the 2D mesh along the axial direction with a separation distance of $\frac16 \Lx^\alpha$, where $\Lx^\alpha$ is the characteristic axial length scale of the pressure introduced in \secref{Ax_var_g} and shown in \figref{3D_fig}(a). The mesh is validated by standard mesh convergence tests~\cite{Ley2017}.

For the 3D modeling of the long capillary tube, we use symmetry considerations to halve the computational domain~\cite{Ley2017}, and a perfectly matched layer (PML) placed at the tube end to suppress acoustic reflections there, see Sec.~II-C of Ref.~\cite{Ley2017}.

This numerical implementation of the model leads to $2\times 10^3$ degrees of freedom (DOF) for the 2D-simulations and $4\times 10^5$ DOF for the 3D simulations. The simulations were performed on a workstation with a 3.5-GHz Intel Xeon CPU E5-1650 v2 dual-core processor, and with a memory of 128~GB RAM.

Finally, we use the $L^2$-norm to numerically compute the relative deviation $\calE(p,p^\mr{ref})$ of a pressure field $p$ from a reference pressure field $p^\mr{ref}$ in the 3D domain $\Omega$ as,
 \beq{rel_dev}
 \calE(p,p^\mr{ref}) = \sqrt{
 \frac{\int_\Omega |p-p^\mr{ref}|^2\: \dm V}{\int_\Omega |p^\mr{ref}|^2\: \dm V}}.
 \eeq
The analogous relative deviation in the 2D domain $\Omegabar$ is called $\bar{\calE}(\bar{p},\bar{p}^\mr{ref})$, where an overbar denote a 2D quantity.

 \begin{table}[t]
 \centering
 \caption{\tablab{params} Parameters used in the numerical simulations of water as the fluid medium at 25 C$^\circ$ \citep{Muller2014}, see also~\secref{numerics3D}.}
 \begin{ruledtabular}
 \begin{tabular}{lccc}
 Parameter & Symbol & Value &  {Unit}  \\ \hline
 Mass density     &   $\rhofl$ & 997.05 & kg~m$^{-3}$ \rule{0ex}{3.0ex} \\
 Compressibility   & $\kapfl$ & 448  &  TPa$^{-1}$  \\
 Speed of sound  & $\cfl$ &1496.7 &  m~s$^{-1}$ \\
 Dynamic viscosity & $\etafl$&  0.890 &  mPa\,s \\
 Bulk viscosity & $\etaBfl$ &2.485 &   mPa\,s \\ \hline
 Actuation displacement & $d_0$ & $0.1$ & nm \rule{0ex}{2ex} \\
 Cross-section length & $\Lc$ & 300 & $\SImum$\\
 Axial domain length & $\Lx^\mr{num}$ & 5.46 & $\mr{cm}$\\
 PML length & $L_\mr{PML}$ & $500$ & $\SIum$\\
 PML strength & $K_\mr{PML}$ & 1000 & --
 \end{tabular}
 \end{ruledtabular}
 \end{table}

\section{The acoustic pressure in a long straight capillary tube of arbitrary cross section} \seclab{theory}

In the following, we calculate the acoustic pressure $p_1(x,y,z)$ satisfying \eqref{p1_gov} to lowest order in the small parameters $k_0\deltas$, \eqref{kdelta}, in a long, straight capillary tube of arbitrary cross section that is invariant in the axial $x$ direction as shown in \figref{3D_fig}(a). Our strategy has two key steps: First, based on our previous analysis of the 2D cross-sectional resonance modes $\bar{p}_1^{mn}(x,y)$ in a rectanglular cross section having integer $m$ and $n$ half-waves in the $y$ and $z$ direction, and including the viscous boundary layer~\cite{Bach2019}, we write an expression for the 2D cross-sectional resonance modes $\bar{p}_1^{\alpha}(x,y)$ in an arbitrary cross section. Second, by using these 2D modes together with the residue theorem, we evaluate the  3D acoustic pressure $p_1(x,y,z)$ satisfying \eqref{p1_gov}  for any frequency $f=\frac{1}{2\pi}\omega$ and actuation function $U_{1\perp}$ as a sum over all resonance modes~$\alpha$.

\subsection{The 2D pressure resonance modes in an arbitrary cross section}
\seclab{theory2Dmode}

In Ref.~\cite{Bach2019}, we studied the special case of a rectangular cross section of side lengths $\Ly$ and $\Lz$. We derived to lowest order in the small parameter $k_0\deltas$, the resonance modes $\pbar^{mn}_1$ with $m$ half-waves in the $y$ direction and $n$ half-waves in the $z$ direction, valid for wave numbers $k_0$ close to the resonance wave number $\bar{k}_0^{mn}$. Here, and in the following, we use the overbar to denote a quantity defined in the cross section $\Omegabar$. With this notation, the expression for $\pbar^{mn}_1$ given in Eq. (12) of Ref.~\cite{Bach2019} becomes
 \bsubal{p1_rectangle}
 \eqlab{p1_rectangle_form}
 \bar{p}_1^{mn}(\kO; y,z)&=\bar{P}_1^{mn} \calGbar^{mn}(k_0) \Rbar^{mn}(y,z),  \quad \text{for $k_0\approx \kbar_0^{mn}$,}
 \\ \eqlab{p1_rectangle_R}
 \Rbar^{mn}(y,z) &= \cos\bigg(\frac{m \pi y}{L_y}\bigg)\cos\bigg(\frac{n \pi z}{L_z} \bigg),
 \\ \eqlab{p1_rectangle_P1}
 \Pbar_1^{mn}&= \frac{\oint_{\dOmegabar} \bar{U}_{1\perp} \Rbar^{mn} \, \dd l}{\int_{\Omegabar}  \big(\Rbar^{mn}\big)^2  \, \dd A },
 \\ \eqlab{p1_rectangle_G}
 \calGbar^{mn}(k_0) &= \frac{ \frac12 \kbar_0^{mn}}{k_0-\kbar_0^{mn}+\frac12 \ii \kbar_0^{mn}  \Gammabar^{mn}}.
 \esubal
The quantities used here have the following meaning: $\Rbar^{mn}(y,z)$ in \eqref{p1_rectangle_R} is the spatial dependency of a given 2D cross-sectional resonance mode.  $\Pbar^{mn}_1$ in \eqref{p1_rectangle_P1} is a coupling coefficient related to the overlap between $\Rbar^{mn}(y,z)$ and the actuation function $\Ubar_{1\perp}(y,z)$ defined in \eqref{p1_gov_U_def} on the boundary $\dOmegabar$ of the cross section $\Omegabar$.  $\calGbar^{mn}(k_0)$ in \eqref{p1_rectangle_G} is the line-shape function of the mode defined in terms of three parameters: the wave number $k_0=\frac{\omega}{\cfl}$, the resonance wave number $\kbar_0^{mn}$, and the minute damping coefficient $\Gammabar^{mn} = \Gamblbar^{mn}+\Gamfl \ll 1$, where the latter is  defined in Eq.~(10) of Ref.~\cite{Bach2019} as the sum of the boundary-layer damping coefficient $\Gamblbar^{mn}$ and the bulk damping coefficient $\Gamfl$ of \eqref{kc_Gamma}.

\subsubsection*{Generalization to an arbitrarily shaped 2D cross section}

To generalize from the rectangular cross section to an arbitrarily shaped cross section, it is helpful to write $\calGbar^{mn}$ as a function not of $\kO$ but of the complex-valued wave number $\kc$ from \eqref{kc_Gamma} for $\kc\approx\kcbar^{mn}$, where  $\kcbar^{mn}$ is the complex-valued resonance wave number. This variable shift is obtained by inserting $\Gammabar^{mn}=\Gamblbar^{mn}+\Gamfl$ in \eqref{p1_rectangle_G},
 \beq{Gbar_rewrite}
 \calGbar^{mn}(\kc) \approx \frac{(\kcbar^{mn})^2}{\kc^2-(\kcbar^{mn})^2} ,
 \quad \kcbar^{mn}
 =\Big(1-\ii  \frac12\Gamblbar^{mn}\Big)\kbar_0^{mn}.
 \eeq
In \eqsref{p1_rectangle}{Gbar_rewrite}, we substitute the mode index $mn$ by $\alpha$, and thereby introduce our main assumption, an expression for the pressure resonance mode $\pbar^\alpha_1$ in an arbitrary cross section valid close to resonance $\kc \approx \kcbar^\alpha$ and to lowest order in the small parameter $k_0 \deltas$,
 \bsubal{p1_general}
 \eqlab{p1_general_form}
 \pbar_1^\alpha(\kc; y,z) &=\bar{P}_1^\alpha \calGbar^\alpha(\kc) \Rbar^\alpha(y,z),  \quad \text{for $\kc \approx \kcbar^\alpha$},
 \\ \eqlab{p1_general_P1}
 \Pbar_1^\alpha &= \frac{\oint_{\dOmegabar} \bar{U}_{1\perp} \Rbar^\alpha \, \dd l}{\int_{\Omegabar}  \big(\Rbar^\alpha\big)^2  \, \dd A },
 \\ \eqlab{p1_general_G}
 \calGbar^\alpha(\kc) &=  \frac{ (\kcbar^\alpha)^2}{\kc^2-(\kcbar^\alpha)^2}, \qquad  \kcbar^{\alpha}
 =\Big(1-\ii \frac12\Gamblbar^{\alpha}\Big)\kbar_0^{\alpha}.
\esubal
Here, the eigenvalue $\kcbar^\alpha$ and eigenfunction $\Rbar^\alpha$ are defined through the 2D eigenvalue problem, corresponding to \eqref{p1_gov}, in the cross section $\Omegabar$ without actuation,
 \begin{figure}[t]
 \includegraphics[width=\columnwidth]{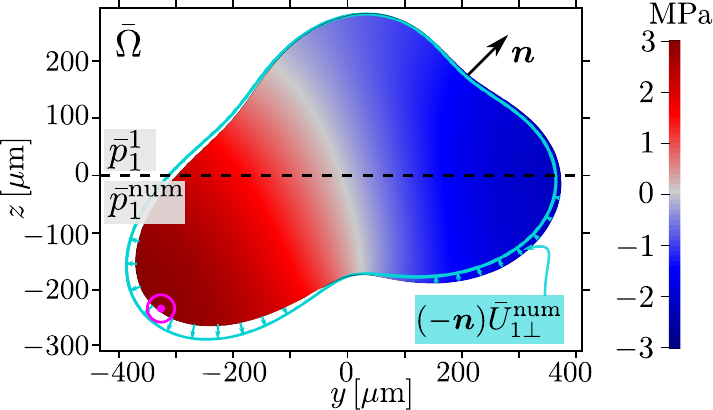}
 \caption{\figlab{validate_2D} Comparison in the cross section $\Omegabar$ (defined in the text of \secref{numeric_impl}) between the 2D pressure mode $\pbar_1^{1}$ from \eqsref{p1_general}{p1_Eigenvalue} with $\alpha = 1$ plotted for $z>0$, and the 2D numerical pressure $\pbar^\mr{num}_1$ from \eqref{p1_gov} plotted for $z<0$, both actuated by the actuation function $\bar{U}_{1\perp}^\mr{num}$ (cyan curve and arrows, see \eqref{Ubarnum}) at the fundamental resonance frequency $\bar{f}^1=1.1341 $~MHz.  The relative deviation defined in \eqref{rel_dev} between $\pbar_1^1$ and $\pbar_1^\mr{num}$ is $\bar{\calE}(\pbar_1^1,\pbar_1^\mr{num}) = 0.14~\%{}$. The encircled magenta point marks the position of the line used in the 3D line plots of \secref{numerics3D}.}
 \end{figure}
 \bsubalat{p1_Eigenvalue}{2} 
 \eqlab{p1_Eigenvalue_helmholtz}
 \lap \Rbar^\alpha+(\kcbar^\alpha)^2 \Rbar^\alpha&=0,
 & \qquad  &\rrr\in\Omegabar,
 \\
 \eqlab{p1_Eigenvalue_bc}
  \calD_\perp \Rbar^\alpha&=0,
 &  & \rrr\in\dOmegabar.
 \esubalat
The resonance frequency $\fbar^\alpha$ of the 2D mode $\alpha$ is found from the real part $\kbar_0^\alpha= \Re(\kcbar^\alpha)$ of the eigenvalue $\kcbar^\alpha$,
 \bal \eqlab{fbar_general}
\fbar^\alpha= \frac{1}{2\pi}\omegabar^\alpha=\frac{1}{2\pi}\cfl \kbar_0^{\alpha}.
\eal
The damping coefficient $\Gammabar^\alpha$ of mode $\alpha$ is written as the sum of the bulk damping coefficient $\Gamfl$ of \eqref{kc_Gamma} and the boundary-layer damping coefficient $\Gamblbar^\alpha$ of \eqref{p1_general_G},
\beq{Gammabar_general}
\Gammabar^\alpha=\Gamblbar^\alpha+\Gamfl, \qquad \Gamblbar^\alpha =-\frac{2\Im\big(\kcbar^\alpha\big)}{\Re\big(\kcbar^\alpha\big)}.
\eeq

\subsubsection*{Numerical validation in the 2D cross section}

 \begin{table}[b]
 \centering
 \caption{\tablab{params_calc} Values for the fundamental mode $\alpha=1$ of the 2D eigenvalue problem obtained by numerical simulation.}
 \begin{ruledtabular}
 \begin{tabular}{lcccc}
 Parameter &  Symbol & Eq. & Value &  {Unit}  \\ \hline
 Eigenvalue   &   $\kcbar^{1}$ & \eqref{p1_Eigenvalue} & $4761.01-3.49 \ii $ & m$^{-1}$ \rule{0ex}{3.0ex} \\
 Eigenfrequency   & $\bar{f}^{1}$& \eqref{fbar_general} & 1.1341  &  MHz \\
 Damping coefficient  & $\Gammabar^{1}$ & \eqref{Gammabar_general}& $0.00148$ & --  \\
 $x$ length scale  & $\Lx^1$ & \eqref{Lx}&  5.46 & mm
 \end{tabular}
 \end{ruledtabular}
 \end{table}
In \figref{validate_2D}, we validate numerically the generalized resonance-mode structure~\eqnoref{p1_general} by using the cross section $\Omegabar$ and the numerical procedure described in \secref{numeric_impl}. We choose the actuation function $\Ubar^\mr{num}_{1\perp}$ along the boundary $\pp\Omegabar$ of $\Omegabar$ to be,
 \beq{Ubarnum}
 \Ubar^\mr{num}_{1\perp}(y,z) = \rhofl \cfl^2 d_0\sin\Big(\frac{\pi}{\Lc} y\Big)\ee^{-\frac{z}{\Lc}}.
 \eeq
We determine numerically the lowest eigenmode $\alpha = 1$ in terms of the eigenfunction $\Rbar^1(y,z)$, \eqref{p1_Eigenvalue}, and eigenfrequency $\fbar^1$, \eqref{fbar_general}, listed in \tabref{params_calc} together with other relevant mode parameters for $\alpha = 1$. Inserting this eigenmode together with $\Ubar^\mr{num}_{1\perp}(y,z)$ and $\kc = \frac{2\pi}{\cO}\big(1+\ii\frac12\Gamfl\big)\:\fbar^1$ into \eqref{p1_general}, we compute the pressure resonance mode $\pbar_1^1(\kc; y,z)$ at the resonance frequency. In \figref{validate_2D} we compare this theoretical result  $\pbar_1^1$ with the direct numerical simulation $\pbar_1^\mr{num}$ obtained from the 2D version of \eqref{p1_gov} at the resonance frequency $\fbar^1$. Qualitatively, we see a smooth transition passing from $\pbar_1^1$ above the dashed line ($z>0$) to $\pbar_1^\mr{num}$ below the dashed line ($z<0$). Quantitatively, the relative difference~\eqnoref{rel_dev} between the semi-analytical $\pbar_1^1$ and the numerical $\pbar_1^\mr{num}$ is found to be $\bar{\calE}(\pbar_1^1,\pbar_1^\mr{num}) = 0.14~\%{}$, which is satisfactory in this approximation to lowest order in the small boundary-layer-width parameter $\kO \deltas=0.24~\%$.

\subsection{The 3D pressure}
\seclab{theory3Dpressure}

Based on expression \eqnoref{p1_general} for the 2D cross-sectional pressure modes $\pbar_1^\alpha$, we now derive the pressure $p_1(x,y,z)$ satisfying \eqref{p1_gov} in the 3D capillary tube. For any given $x$-dependent function $\phi(x)$, we denote its Fourier transform by $\hat\phi(\kx)$, see \appref{Fourier}. The 3D pressure is calculated from the inverse Fourier transform,
 \bsub
 \bal \eqlab{p1_xyz}
 p_1(x,y,z)&=\int_{-\infty}^\infty \hat{p}_1(\kx;y,z)\: \ee^{\ii \kx x}\, \frac{\dd \kx}{2\pi}.
 \eal
Since the integrand $\hat{p}_1(\kx;y,z)\:\ee^{\ii \kx x}$ is a function of the complex-valued wave number $\kx$, we evaluate the integral using the residue theorem for an appropriate closed contour $\gamma$ in the complex $\kx$-plane,
% ZZZ leading to the expression,
and find
 \bal \eqlab{p1_xyz_Res}
 &p_1(x,y,z)=\sum_{\kx^{\alpha}\; \mr{inside}\; \gamma} \ii \: \mr{Res}\big(\hat p_1(\kx;y,z) \ee^{\ii \kx x},\kx^{\alpha}\big),
 \eal
 \esub
summing over the residues $\mr{Res}\big(\hat p_1(\kx;y,z) \ee^{\ii \kx x},\kx^{\alpha}\big)$ of all poles $\kx^{\alpha}$ inside the closed contour $\gamma$. To obtain these residues, we only need an expression for $\hat p_1(\kx;y,z) \ee^{\ii \kx x}$ valid close to  $\kx^\alpha$. The Fourier transform $\hat{p}_1(\kx;y,z)$ satisfies the Fourier-transformed Helmholtz problem \eqnoref{p1_gov},
 \bsubal{p1_Fourier}
 &\lap \hat{p}_1(\kx;y,z)+\big(\kc^2-\kx^2\big) \hat{p}_1(\kx;y,z)=0, \: \rrr \in \Omegabar,
 \\
 &\calD_\perp \hat{p}_1(\kx;y,z) = k_0^2\hat{U}_{1\perp}(\kx;y,z),\: \rrr\in\pp \Omegabar,
 \esubal
where $\pp\Omegabar$ is the boundary of $\Omegabar$. We note that \eqref{p1_Fourier} for $\phat_1(\kx;y,z)$ is similar to \eqref{p1_gov} for an $x$-independent pressure $p_1 = \pbar_1(y,z)$ with the substitutions $\kc^2 \rightarrow \kc^2-\kx^2$ and $ \Ubar_{1\perp}(y,z)\rightarrow \hat{U}_{1\perp}(\kx;y,z)$, see \appref{app_calD} for details. Using these substitutions in \eqref{p1_general}, we obtain the result for $\hat{p}_1(\kx;y,z)\ee^{\ii \kx x}$ valid for $\kc^2-\kx^2 \approx (\kcbar^\alpha)^2$ and to lowest order in the small parameter $k_0\deltas$,
 \bsubal{phat_exp}
 \eqlab{phat_int}
 \hat p_1(\kx;y,z) \ee^{\ii \kx x} &\approx
 \frac{ -(\kcbar^\alpha)^2 \Rbar^\alpha \ee^{\ii \kx x}}{(\kx)^2-(\kx^ \alpha)^2}\, \frac{\int_{\dOmegabar} \hat{U}_{1\perp}(\kx;y,z) \Rbar^\alpha\, \dd l}{\int_{\Omegabar} \big(\Rbar^\alpha\big)^2  \, \dd A },
 \\
 \eqlab{kx_alpha}
 \kx^\alpha&=\sqrt{\kc^2-(\kcbar^\alpha)^2}.
 \esubal
From \eqref{phat_int} we see that $\hat{p}_1(y,z,\kx)\ee^{\ii \kx x}$ has simple poles in the complex $\kx$-plane at 
% ZZZ the values 
$\kx=\pm \kx^\alpha$, and therefore the residues $\mr{Res}\big(\hat p_1(\kx;y,z) \ee^{\ii \kx x},\kx^{\alpha}\big)$ used in the sum \eqnoref{p1_xyz_Res} can be found analytically
% ZZZ analytically. In \appref{Res_details} we describe the details in applying the residue theorem in
analytically, see \appref{Res_details}.
The resulting expression for $p_1(x,y,z)$, valid for \emph{all} frequencies and to lowest order in $k_0\deltas$, is
 \bsubal{p1_3d_sol}
 \eqlab{p1_3d_sum}
 p_1(x,y,z)&= \sum_\alpha p_1^\alpha(x,y,z),
 \\ \eqlab{p1_3d_alpha}
 p_1^\alpha(x,y,z)&= P_1^\alpha(x) \calGbar^\alpha(\kc) \Rbar^\alpha(y,z),
 \\ \eqlab{p1_3d_P1}
 P^\alpha_1(x)&= \Bigg[\frac{\int_{\dOmegabar} U_{1\perp}\Rbar^\alpha\, \dd l}{\int_{\Omegabar} \big(\Rbar^\alpha \big)^2  \, \dd A } * g^\alpha \Bigg](x),
 \\  \eqlab{p1_3d_g}
 g^\alpha(x) &= \frac{-\ii \kx^\alpha}{2} \ee^{\ii\kx^\alpha |x|}.
 \esubal
Here, the asterisk \qmarks{$*$} denotes the usual functional convolution in the $x$ coordinate, see \eqref{convolution}. $g^\alpha(x)$ is the Green's function in the axial direction of mode $\alpha$ corresponding to a delta-function actuation at $x=0$, given by $U_{1\perp}(x,y,z)=\Ubar_{1\perp}(y,z) \Lact \delta(x)$, as this actuation yields
 \bal \eqlab{chidelta}
 P_1^\alpha(x)=\Pbar_1^\alpha \Lact g^\alpha(x),
 \eal
where $\Lact$  is an  actuation strength of dimension length, and $\Pbar_1^\alpha$ is the 2D coupling coefficient defined in \eqref{p1_rectangle_P1}.

 \begin{figure}[t]
 \includegraphics[width=\columnwidth]{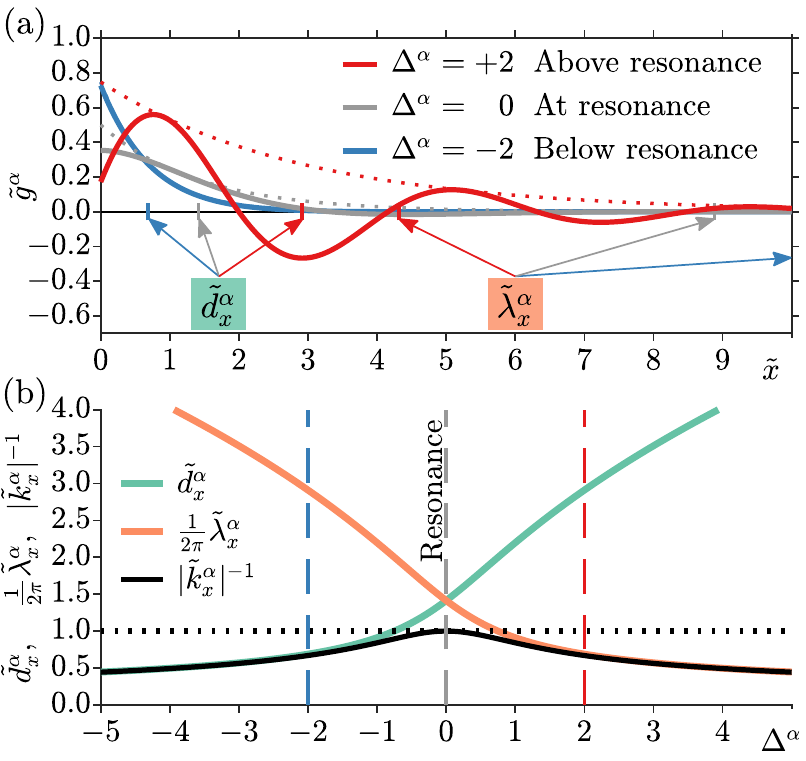}
 \caption{\figlab{g_lengthscales} Plots of the rescaled quantities of \eqref{tildeLengths}.
 (a) The complex-valued Green's function $\tilde{g}^\alpha(x)$ of \eqref{p1_3d_g} plotted as the real part (solid) and modulus (dotted) for three values of $\Delta^\alpha$ of \eqref{kx_size_Gamma}: Below ($-2$), at (0), and above (2) the $\alpha$ resonance. (b) The decay length $d_x^\alpha$ and wave length $\lambda_x^\alpha$, \eqref{dx_lambdax}, as well as $\big|k_x^\alpha\big|^{-1}$, \eqref{kx_size_Gamma}, plotted versus $\Delta^\alpha$. The three vertical dashed lines mark the values used in (a).}
 \end{figure}

\subsection{The axial length scales of each mode} \seclab{Ax_var_g}

The axial dependency of the pressure $p_1(x,y,z)$ is given in \eqsref{p1_3d_P1}{p1_3d_g} by the actuation function $U_{1\perp}$ and the Green's function $g^\alpha$. The latter
leads to three axial length scales that characterize each  mode $\alpha$. First, by using $\kc^2=(1+\ii\Gamfl) k_0^2$, \eqref{kc_Gamma}, and $(\kcbar^{\alpha})^2=(1-\ii\Gamblbar^\alpha) (\kbar_0^\alpha)^2$, \eqref{p1_general_G}, as well as the assumption $\Gammabar^\alpha \ll1$, we write the $x$ wave number $\kx^\alpha$ in \eqref{kx_alpha} of mode $\alpha$ as
 \beq{kx_size_Gamma}
 \kx^{\alpha} = \kcbar^\alpha \frac{1}{\sqrt{\calGbar^{\alpha}}}
 \approx  \kbar_0^\alpha \sqrt{\Gammabar^\alpha} \sqrt{\Delta^\alpha+\ii },
 \qquad  \Delta^\alpha =\frac{k_0^2-(\kbar_0^\alpha)^2}{(\kbar_0^\alpha)^2  \Gammabar^\alpha}.
 \eeq
Here, $\Delta^\alpha$ is the difference between the square of the wave number $k_0=\frac{\omega}{\cfl}$ and the
resonance wave number $\kbar_0^\alpha$, \eqref{fbar_general}, of the mode $\alpha$
scaled by $(\kbar_0^\alpha)^2  \Gammabar^\alpha$. From \eqref{kx_size_Gamma}, we identify the characteristic length scale $\Lx^\alpha$ of variation of the mode $\alpha$ in the $x$ direction as,
 \beq{Lx}
 \Lx^\alpha = \frac{1}{\kbar_0^\alpha\sqrt{ \Gammabar^\alpha}}.
 \eeq
Then, by using \eqsref{kx_size_Gamma}{Lx}, we write the decay length $d_x^\alpha$ and wave length $\lambda_x^\alpha$ of the function $g^\alpha(x)$ in \eqref{p1_3d_g} as,
 \bsubalat{dx_lambdax}{2}
 \eqlab{dx}
 d_x^\alpha&=\frac{1}{\Im(\kx^{\alpha})} & &= \Lx^\alpha \frac{1}{\Im(\sqrt{\Delta^\alpha +\ii})},
 \\ \eqlab{lambdax}
 \lambda_x^\alpha&=2\pi\frac{1}{\Re(\kx^{\alpha})} & &= \Lx^\alpha \frac{2\pi}{\Re(\sqrt{\Delta^\alpha +\ii})}.
 \esubalat
In the following, a tilde is used to denote rescaling by $\Lx^\alpha$ in the axial direction,
 \bsubalat{tildeLengths}{3}
 \xti &= \frac{x}{L_x^\alpha}, &\qquad
 \tilde{d}_x^\alpha &= \frac{d_x^\alpha}{\Lx^\alpha},  & \qquad
 \tilde{\lambda}_x^\alpha &= \frac{\lambda_x^\alpha}{\Lx^\alpha}, \\
 \tilde{k}_x^\alpha &= \Lx^\alpha \kx^\alpha, & \qquad
 \tilde{g}^\alpha &= \Lx^\alpha g^\alpha.
 \esubalat
In \figref{g_lengthscales}(a), we plot the rescaled Green's function $\tilde{g}^\alpha(x)$ for three different frequencies: above resonance ($k_0=\kbar_0^\alpha +\kbar_0^\alpha \Gammabar^\alpha$, $\Delta^\alpha = 2$), where it is propagating, at resonance ($k_0=\kbar_0^\alpha$, $\Delta^\alpha = 0$), and below resonance ($k_0=\kbar_0^\alpha -\kbar_0^\alpha \Gammabar^\alpha$, $\Delta^\alpha = -2$), where it is  evanescent. In \figref{g_lengthscales}(b), we plot the decay length $d_x^\alpha$ and the wave length $\lambda_x^\alpha$ of \eqref{dx_lambdax} as a function of the actuation frequency for frequencies close to resonance $k_0\approx \kbar_0^\alpha$. For frequencies just below resonance, $d_x^\alpha$ is small and $\lambda_x^\alpha$ is large, and vice versa for frequencies just above the resonance frequency. The three values of $\Delta^\alpha$ used in  \figref{g_lengthscales}(a) are marked by dashed vertical lines of the same color in \figref{g_lengthscales}(b).

As an example of the characteristic length scale of the axial pressure variation $\Lx^\alpha$, we consider a standing vertical half wave in a rectangular cross section of height $\Lz$ with the wave number $\kbar_0^\alpha \approx \frac{\pi}{\Lz}$, for which  \eqref{Lx} leads to the estimate $\Lx^\alpha \approx \frac{1}{\pi \sqrt{\Gammabar^{\alpha}}}\Lz$. For realistic values of the damping coefficient, $0.001 < \Gammabar^{\alpha} < 0.01$ \cite{Hahn2015}, we obtain $10 \Lz > L^\alpha_x > 3 \Lz$.

\section{The axial dependency of the pressure for a separable actuation}\seclab{envelope_calc}

Above, we described the characteristic length scales of the Green's function $g^\alpha(x)$ entering in \eqref{p1_3d_P1} for the pressure amplitude $P^\alpha_1(x)$. Now, we calculate the axial variation of the pressure mode $p_1^\alpha$ for a given separable model-actuation $U_{1\perp}$ having the dimensionless axial dependency  $\psiact(x)$,
 \beq{U_sep}
 U_{1\perp}(x,y,z)=\Ubar_{1\perp}(y,z)\:\psiact(x).
 \eeq
In this case, the 3D mode $p_1^\alpha(x,y,z)$ in \eqref{p1_3d_alpha} becomes a product of the 2D mode $\pbar_1^\alpha(y,z)$ and the dimensionless axial dependency $\chi^\alpha(x)$ of the pressure,
 \bsubal{P1_sep}
 p_1^\alpha(x,y,z)&=\pbar_1^\alpha(y,z)  \chi^\alpha(x),
 \\ \eqlab{chi_sep}
 \chi^\alpha(x) &=[\psiact* g^\alpha](x),
 \esubal
which by \eqref{p1_3d_sum} leads to the pressure $p_1(x,y,z)$,
 \beq{p1_3D_sep}
  p_1(x,y,z) = \sum_\alpha  \pbar_1^\alpha(y,z)  \chi^\alpha(x),
  \quad \text{for any frequency $f$}.
 \eeq
When actuating the system near one of the 2D resonances, say $\alpha = \alpha'$, we obtain the simplified expression
 \beq{p1_3D_sep_approx}
  p_1(x,y,z) \approx \pbar_1^{\alpha'}(y,z)  \chi^{\alpha'}(x),
  \quad \text{for }\; f \approx \fbar^{\alpha'}.
 \eeq

\subsection{A box actuation with sharp steps}

In capillary-tube devices used for acoustic trapping, a piezoelectric transducer is usually placed below the capillary tube in a confined region\cite{Hammarstrom2010, Hammarstrom2012, Carugo2011, Mishra2014, Lei2013, Evander2015}. To mimic such an actuation, we consider the box-shaped  axial dependency $\psiactbox (x)$ of the actuation, which is unity in the actuation region of length $\Lact$ and sharply steps down to zero outside this region, as sketched in \figref{boxact},
 \bsub
 \bal\eqlab{psibox}
 \psiactbox(x)= \begin{cases}
 1, \qquad  & |x|<\frac12 \Lact, \\
 0, & |x|>\frac12 \Lact.
 \end{cases}
 \eal
Using this actuation in the convolution \eqnoref{chi_sep} yields the axial dependency $\chibox^\alpha(x)$ of the pressure,
 \begin{figure}[t]
 \includegraphics[width=\columnwidth]{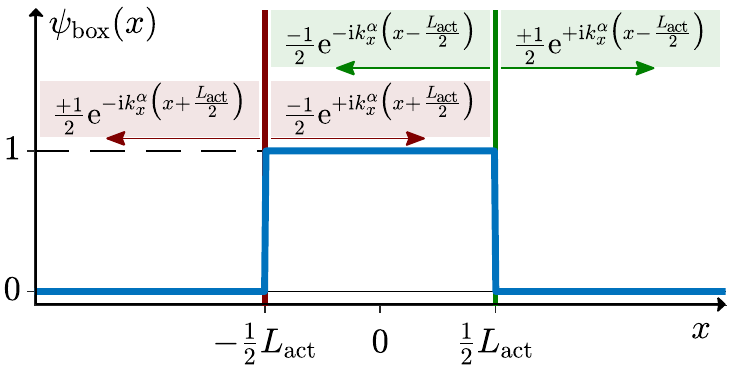}
 \caption{\figlab{boxact} The box-shaped axial dependency $\psiactbox(x)$ (blue) of the model actuation of width $\Lact$, as well as the resulting traveling-wave components excited by the left step at $x=-\frac12 \Lact$ (red) and by the right step at $x=\frac12 \Lact$ (green).}
 \end{figure}
 \bal\eqlab{chibox}
 \chibox^\alpha(x)=\begin{cases}
1-\ee^{\ii\kx^{\alpha}\frac{\Lact}{2}}\cos\big(\kx^\alpha x\big),
 \quad &|x| <\frac12 \Lact,
 \\
 -\ii\sin\Big(\kx^{\alpha}\frac{\Lact}{2}\Big)\ee^{\ii\kx^{\alpha} |x|},
 \quad &|x| >\frac12 \Lact.
 \end{cases}
 \eal
 \esub
By writing the sine and cosine factors in terms of travelling waves, it is found that two waves travel away from each step at $x=\pm\frac12 \Lact$ as sketched in \figref{boxact}.

In the limit $\Lact \ll \Lx^\alpha$, where the actuation is confined to a region much narrower than the axial pressure length scale $\Lx^\alpha$ of \eqref{Lx}, the axial dependency of the pressure is well approximated by a simplified expression $\chidelta^{\alpha}(x)$ from \eqref{chidelta},
 \beq{chidelta2}
 \chibox^{\alpha} (x)\approx \chidelta^{\alpha}(x)= \Lact g^\alpha(x), \quad \text{for }\Lact \ll \Lx^\alpha,
 \eeq
where $g^\alpha(x)$ is defined in \eqref{p1_3d_g} and plotted in \figref{g_lengthscales}(a) for different frequencies.

\subsection{A box actuation with smooth steps}

In the following numerical validation, we consider the more realistic box-actuation function $U_{1\perp}^\mr{num}(x,y,z)=\Ubar_{1\perp}^\mr{num}(y,z)\psiactnum(x)$, which separates as \eqref{U_sep} with an axial dependency  $\psiactnum(x)$, similar to the box shape $\psiactbox(x)$ in \eqref{psibox}, but which has smooth transistions of width $\dact$ at the steps $x=\pm \frac12 \Lact$,
 \bsubal{actuation_num}
 U_{1\perp}^\mr{num}(x,y,z)&=\Ubar_{1\perp}^\mr{num}(y,z)\psiactnum(x),
 \quad \text{at } \pp\Omega,
 \\ \eqlab{psiactnum}
 \psiactnum(x)&= \frac{1}{1+\ee^{\frac{4(x-\frac12\Lact)}{\dact}}}
 - \frac{1}{1+\ee^{\frac{4(x+\frac12\Lact)}{\dact}}}.
 \esubal
Here, $\Ubar_{1\perp}^\mr{num}$ is defined in \eqref{Ubarnum} and shown in \figref{validate_2D}.

\subsection{Numerical validation of the 3D mode method} \seclab{numerics3D}

Using COMSOL Multiphysics as described in \secref{numeric_impl}, we validate the semi-analytical expression~\eqnoref{p1_3D_sep_approx} for $p_1(x,y,z)$ by the direct numerical solution $p_1^\mr{num}$ of \eqref{p1_gov} in the capillary tube $\Omega$ sketched in \figref{3D_fig}, both actuated at the resonance frequency $f=\fbar^{1}$ of the fundamental mode $\alpha=1$ of the 2D cross section $\Omegabar$. We assume that the tube is mirror-symmetric around the $y$-$z$ plane through $x=0$, and take the length of the computational domain to be $L_x^\mathrm{num} = 10\Lx^1=5.46\, \SIcm$, see \tabref{params_calc}. At $x=0$ we impose the symmetry boundary condition $\pp_x p_1^\mathrm{num}=0$, and at $x=L_x^\mathrm{num}$ we place the perfectly matched layer (PML) mentioned in \secref{numeric_impl} to remove pressure wave reflections from the tube end. The wavy boundary $\pp\Omegabar$ of the cross section $\Omegabar$ is defined in \secref{numeric_impl}.

Using the 2D pressure mode~\eqnoref{p1_general} $\bar{p}_1^1(y,z)$ obtained from the eigenvalue problem~\eqnoref{p1_Eigenvalue}, we construct the 3D pressure mode~\eqnoref{P1_sep}, as $p_1^1(x,y,z)=\pbar_1^1(y,z)  \chi^1(x)$. In the following, we use the analytically known axial dependencies $\chibox^1(x)$, \eqref{chibox}, and $\chidelta^1$, \eqref{chidelta},  of the pressure to estimate the pressure obtained numerically from the actuation profile $\psiactnum(x)$ in \eqref{psiactnum}.

To quantify the numerical validation, we compute at the resonance $f = \fbar^1$ the relative deviation $\calE(p_1^1,p_1^\mr{num})$ defined in \eqref{rel_dev} between the semi-analytical 3D pressure mode $p_1^1$ with the box actuation $\psiactbox(x)$ of \eqref{psibox} and the direct numerical 3D pressure $p_1^\mr{num}$ with the  smoothen-box actuation $\psiactnum(x)$ of \eqref{psiactnum},
 \bsubalat{p11p1num}{2}
 \eqlab{p11_validation}
 p_1^1 &=  \bar{p}_1^1(y,z)\chi^1_\mr{box}(x),&& \text{ with } \psiactbox(x) \text{ at } \fbar^1,
 \\
 \eqlab{p1num_validation}
 p_1^\mr{num} &= p_1(x,y,z),&& \text{ with } \psiactnum(x) \text{ at } \fbar^1\!.
 \esubalat

In \figref{validate_d}, we study the axial dependency of the pressure for varying actuation step width $\dact$ and fixed actuation length of $\Lact=2\Lx^1$, rescaled as in \eqref{tildeLengths} by the characteristic length scale $\Lx^1$,
 \beq{dactti_Lactti_xti}
 \dactti=\frac{\dact}{\Lx^1}, \qquad \Lactti=\frac{\Lact}{\Lx^1}.
 \eeq
In \figref{validate_d}(b) is shown that for small $\dact$ the semi-analytical expression \eqnoref{p11_validation} is a good approximation for all $\xti$. For large $\dact$ it deviates significantly from the numerical solution \eqnoref{p1num_validation} inside the actuation region for $|\xti| <1$, whereas it remains a good approximation outside for $|\xti| <1$. This is quantified in the inset of \figref{validate_d}(b), showing that the deviation $\calE_\mathrm{box}$ (solid line) is around 1~\% for a sufficiently narrow actuation step width $\dactti \lesssim 0.4$.

In \figref{validate_L} we vary the actuation length $\Lact$ and keep the actuation step width fixed at $\dactti=0.1$. For all actuation lengths $\Lactti$, the semi-analytical expression \eqnoref{p11_validation} $\pbar_1^1\chibox^1$ (dashed magenta lines) approximates well the full numerical solution $p_1^\mr{num}$ \eqnoref{p1num_validation} (solid green lines), whereas  $\bar{p}_1^1(y,z)\chidelta^1(x)$ (black dotted lines), see \eqref{chidelta}, as expected is only a good approximation in the narrow-actuation limit $\Lactti \ll 1$. In the inset of \figref{validate_L}(b) the relative deviations of these approximations from $p_1^\mr{num}$ are quantified by $\calE_\mathrm{box}$ (solid line) and $\calE_\delta$ (dotted line).

 \begin{figure}[t]
 \includegraphics[width=\columnwidth]{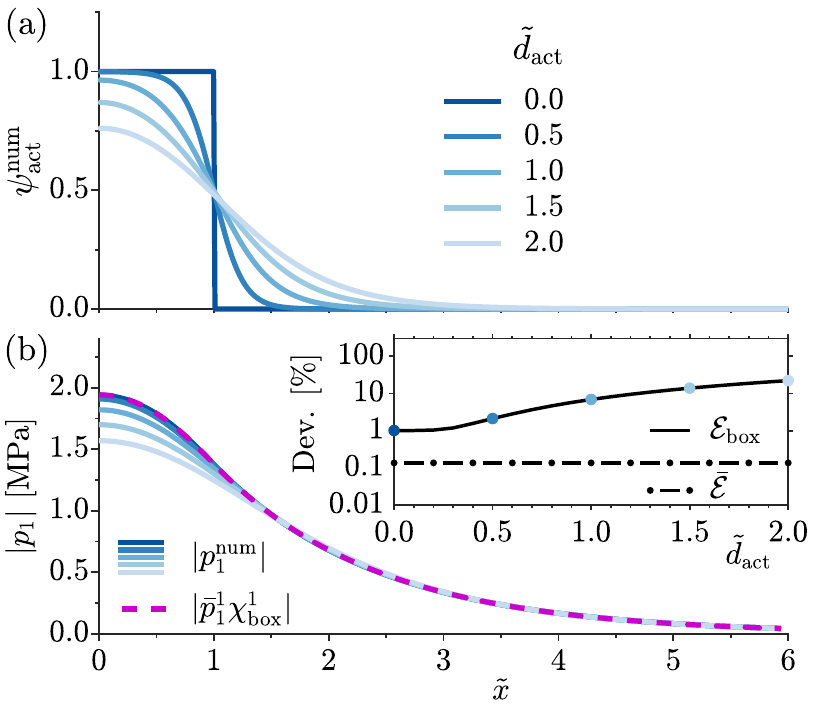}
\caption{\figlab{validate_d} The acoustic pressure for varying actuation step width $\dactti$ and fixed actuation length $\Lactti=2$. (a) The actuation profile $\psiactnum(x)$ used in the full numerical simulation for $\dactti=0.0$ (dark blue) to $\dactti=2.0$ (light blue). (b) Line plots of the magnitude of the acoustic pressure along the axis parallel to the $x$-axis shown in \figref{validate_2D}, with  $|p_1^1| = |\pbar_1^1\chibox^1|$ (magenta dashed line) from \eqref{p11_validation}, and $|p_1^\mr{num}|$ (blue lines) from \eqref{p1num_validation} obtained from the 3D simulation by using the actuation of same color shown in (a). The inset shows the relative deviation~\eqnoref{rel_dev} $\calE_\mr{box}$ (solid line) of the pressure mode $\bar{p}_1^1(y,z) \chibox^1(x)$ from the numerical pressure $p_1^\mr{num}$, as well as the deviation for the 2D calculation given in \secref{theory2Dmode}, $\bar{\calE} = 0.14~\%$ (dot-dashed line).}
\end{figure}

 \begin{figure}[t]
 \includegraphics[width=\columnwidth]{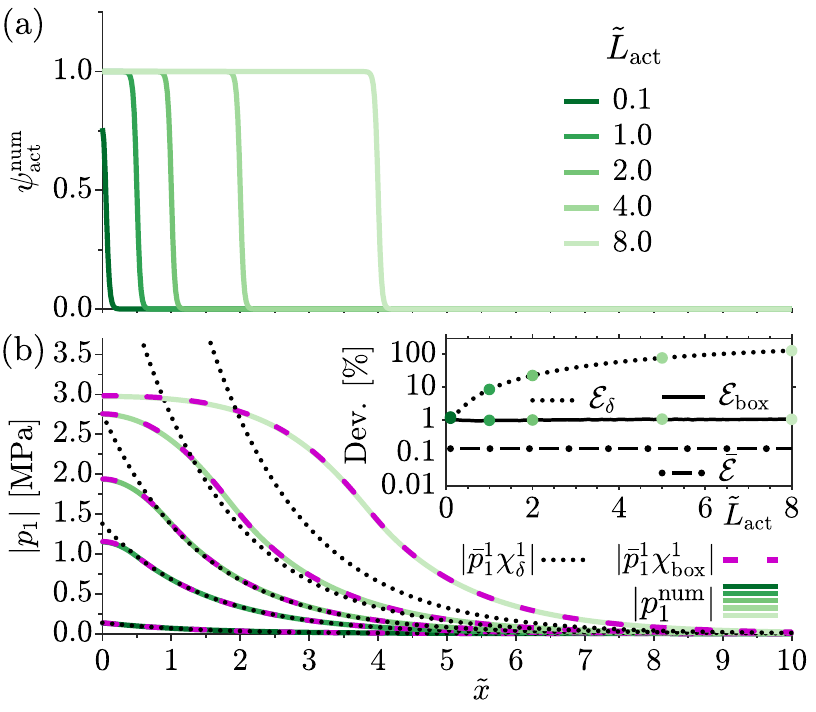}
 \caption{\figlab{validate_L} The acoustic pressure for varying actuation length $\Lactti$ and fixed actuation step width $\dactti=0.1$. (a) The actuation profile $\psiactnum(x)$ used in the full numerical simulations for $\Lactti=0.1$ (dark green) to $\Lactti=8.0$ (light green). (b) Line plots of the magnitude of the acoustic pressure along the axis parallel to the $x$ axis shown in \figref{validate_2D}. The green graphs show the pressure $p_1^\mr{num}$ obtained from the 3D simulation by using the actuation of same color shown in (a). The dashed magenta lines show $p_1^1=\bar{p}_1^1\chibox^1$ from \eqref{p1num_validation}, and the dotted black lines show the pressure $\bar{p}_1^1\chidelta^1$ valid in limit $\Lactti \ll 1$, see \eqref{chidelta}. The inset shows the deviation $\calE$ from to the reference pressure $p_1^\mr{num}$ \eqref{rel_dev}, for $\bar{p}_1^1\chibox^1$ ($\calE_\mr{box}$, solid), for $\bar{p}_1^1\chi^1_\delta$ ($\calE_\delta$, dotted), as well as the deviation for the 2D calculation given in \secref{theory2Dmode}, $\bar{\calE} = 0.14~\%$ (dot-dashed line).}
 \end{figure}

\section{Physical time-averaged quantities close to resonance}\seclab{physical}

In typical experiments on acoustofluidic devices, the MHz oscillation of the acoustic pressure $p_1$ is not observed directly. We therefore study the physical time-averaged quantities given in \eqref{Energies_general}.

\subsection{Time-averaged quantities for a single mode}

We study a single-mode pressure resonance of the form $p_1\approx p_1^\alpha=\pbar_1^{\alpha}(y,z)\chi^\alpha(x)$, see \eqref{p1_3D_sep_approx}. Inserting this form in \eqref{Energies_general} together with the rescaled axial coordinate $\xti = x/\Lx^\alpha$ and the corresponding derivative $\pp_\xti =\Lx^\alpha\:\pp_x$, both scaled with the characteristic axial length scale $\Lx^\alpha$ from \eqref{Lx}, we obtain the time-averaged quantities,
 \bsubal{time-averaged_quantities}
 \eqlab{Epot_alpha}
 \Epot^\alpha&=\Epotbar^\alpha \big|\chi^\alpha \big|^2,
 \\ \eqlab{Ekin_alpha}
 \Ekin^\alpha &=\Ekinbar^\alpha \big|\chi^\alpha\big|^2+ \frac{\Epotbar^{\alpha}}{(k_0\Lx^\alpha)^2} \big|\pp_\xti \chi^\alpha\big|^2  \approx \Ekinbar^\alpha \big|\chi^\alpha\big|^2,
 \\ \eqlab{Eac_alpha}
 \Eac^\alpha &\approx \Eacbar^\alpha \big|\chi^\alpha\big|^2,
 \\ \eqlab{Urad_alpha}
 \Urad^\alpha &\approx   \Uradbar^\alpha \big|\chi^\alpha\big|^2,
 \\ \eqlab{Frad_alpha}
 \FFFrad^\alpha &\approx \FFFradbar^\alpha \big|\chi^\alpha\big|^2-\sqrt{\Gammabar^\alpha}\kbar_0^\alpha \Uradbar^{\alpha} \pp_\xti \big|\chi^\alpha\big|^2 \ex,
 \\ \eqlab{Sac_alpha}
 \SSSac^\alpha &=\SSSacbar^\alpha \big|\chi^\alpha\big|^2 +\sqrt{\Gammabar^\alpha}\cfl\Epotbar^\alpha  \Im\big[2(\chi^\alpha)^* \pp_\xti \chi^\alpha\big] \ex.
 \esubal
Here, the overbar denote a cross-section quantity obtained by using the cross-section resonance pressure $\pbar^\alpha_1(y,z)$ in \eqref{Energies_general}. In Eqs. \eqnoref{Ekin_alpha}-\eqnoref{Frad_alpha} we used that $(k_0 L_x^\alpha)^{-2} \approx \Gammabar^{\alpha}$, see \eqref{Lx}, which is here assumed to be much smaller than unity. We note that the time-averaged quantities listed in \eqref{time-averaged_quantities} have three different axial dependencies: $|\chi^\alpha|^2$ which is the axial dependency of $\Ekin^\alpha$, $\Epot^\alpha$, $\Eac^\alpha$, $\Urad^\alpha$, $(\FFFrad^\alpha)_{yz}$, and $(\SSSac^\alpha)_{yz}$. $\pp_\xti |\chi^\alpha|^2$ which is the axial dependency of the axial radiation force $F^\alpha_{\mr{rad},x}$. And finally $\Im\big[2(\chi^\alpha)^* \pp_\xti \chi^\alpha\big]$ which is the axial dependency of the axial energy flux density $S^\alpha_{\mr{ac},x}$.

\subsection{Time-averaged quantities for the box actuation}

In \figref{contours}, we use the box actuation with $\chi^\alpha=\chibox^\alpha$, see \eqref{chibox}, to make contour plots in the $\Lactti$-$\xti$ plane of the axial dependency of the time-averaged quantities listed in \eqref{time-averaged_quantities}. For each quantity, we choose the rescaled frequency $\Delta^\alpha\approx \frac{f_0-f_0^\alpha}{\frac12 f_0 \Gammabar^\alpha}$, see \eqref{kx_size_Gamma}, to obtain the largest possible value of that physical quantity, \ie\ to optimize $|\calGbar^\alpha|^2 |\chibox^\alpha|^2$, $|\calGbar^\alpha|^2 \pp_\xti |\chibox^\alpha|^2$, and $|\calGbar^\alpha|^2 \Im\big[2(\chibox^\alpha)^* \pp_\xti \chibox^\alpha\big]$, respectively.

In \figref{contours}(a) is shown a contour plot of the axial dependency $|\chibox^\alpha (x)|^2$ of the acoustic energy density $\Eac^\alpha$, \eqref{Eac_alpha}, and cross-sectional radiation force $(\FFFrad^\alpha)_{y,z}$, \eqref{Frad_alpha}. The blue contour delimit the region where the 3D acoustic energy density $\Eac^\alpha$ is larger than the 2D acoustic energy density $\Eacbar^\alpha$. The orange dot marks the maximum obtainable acoustic energy density which is $\max\{\Eac^\alpha\}=1.23 \Eacbar^\alpha$, found at the optimal actuation length $\Lactti \approx 6.1$.

In \figref{contours}(b) is shown a contour plot of the axial dependency $\pp_\xti |\chibox^\alpha (x)|^2$ of the axial acoustic radiation force $F_{\mathrm{rad},x}^\alpha$, see \eqref{Frad_alpha}. The orange dots mark the maximum obtainable axial trapping force which is found to be $\mr{max}\{F_{\mathrm{rad},x}^\alpha\} =\mp 0.63 \sqrt{\Gammabar^\alpha} \kbar_0^\alpha \Uradbar^\alpha$, for the optimal actuation length $\Lactti\approx 4.1$. For this optimal value, the force is largest at the axial position $\xti=\pm 1.3$, which is around 17\% inside the actuation region.

Finally, in \figref{contours}(c) is shown a contour plot of the axial dependency $\Im\big[2(\chibox^\alpha)^* \pp_\xti \chibox^\alpha\big]$ of the axial acoustic energy flux density $S_{\mathrm{ac},x}^\alpha$, see \eqref{Sac_alpha}. Clearly, the energy is always transported away from the actuation region at the edges of the actuation domain $\xti \approx \pm \frac12 \Lactti$. The orange dots mark the largest obtainable axial energy flux density which is $\mr{max}\{S_{\mathrm{ac},x}^\alpha\} = \pm 0.66 \sqrt{\Gammabar^\alpha} \cfl \Epotbar^\alpha$.

 \begin{figure}[t]
 \includegraphics[width=\columnwidth]{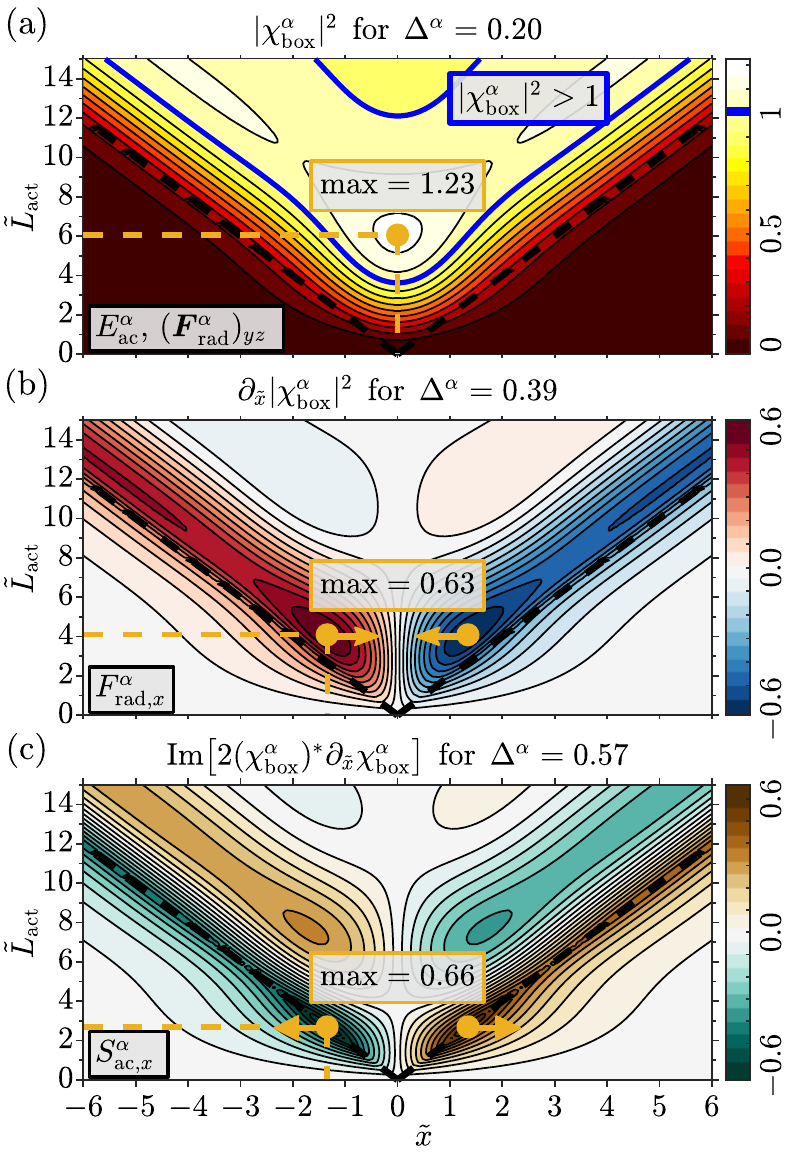}
 \caption{\figlab{contours} The axial dependency of the time-averaged physical quantities of \eqref{time-averaged_quantities} obtained by using $\chibox^\alpha$ from \eqref{chibox} as the axial dependency of the pressure. In each plot, the rescaled frequency  $\Delta^\alpha\approx \frac{f_0-f_0^\alpha}{\frac12 f_0 \Gammabar^\alpha}$, \eqref{kx_size_Gamma}, is chosen to maximize the corresponding physical quantity, and the maximum value is marked by yellow points. The black dashed lines show the actuation edge $\xti=\pm \frac12 \Lactti$. (a) The axial dependency $|\chi_\mr{box}^\alpha|^2$ of the acoustic energy density $\Eac^\alpha$, and the cross-sectional radiation force $(\FFFrad)_{yz}$, where the contours are separated by 0.1 and the blue lines mark the area where the axial dependency of $|\chi_\mr{box}^\alpha|^2$ exceeds unity. (b) The axial dependency $\pp_{\xti}|\chibox^\alpha|^2$ of the axial radiation force $F_{\mathrm{rad},x}^\alpha$. (c) The axial dependency $\Im[2(\chibox^\alpha)^*\pp_{\xti} \chibox^\alpha]$ of the axial energy flux density $S_{\mathrm{ac},x}^\alpha$.}
\end{figure}

\subsection{Example: a standing half-wave resonance in a rectangular cross section}\seclab{halfwave}
A standard device for acoustic trapping is the capillary tube with the rectangular cross section $0<y<\Ly$ and $0<z<\Lz$, where a standing-half-wave resonance in the vertical  $z$ direction is excited \cite{Hammarstrom2010, Hammarstrom2012, Carugo2011, Mishra2014, Lei2013, Evander2015}. Using \eqref{p1_rectangle} with $\kbar_0^{01}=\frac{\pi}{\Lz}$ for the 2D pressure mode $\pbar_1^{01}=\Pbar^{01}_1 \calGbar^{01} \cos\big(\kbar_0^{01} z\big)$, we evaluate the  cross-sectional radiation force $F_{\mr{rad},z}^{01}(x,z)$ from \eqref{Frad_alpha} as,
 \bsub
 \eqlab{Frad_rect}
 \bal \eqlab{Frad_z_rect}
 \frac{F_{\mr{rad},z}^{01}(x,z)}{4\pi a^3 \kbar_0^{01}  \avr{\Eacbar^{01}}} \approx  \Phi       \sin\big(2 \kbar_0^{01} z\big)  \big|\chi^{01}(x)\big|^2,
 \eal
where $\Phi=\frac13 f_0+\frac12 f_1$ is the usual acoustic contrast factor~\cite{Yosioka1955, Settnes2012, Karlsen2015}, and $\avr{\Eacbar^{01}} = \frac14 \kapfl \big|\Pbar^{01} \calGbar^{01}\big|^2$ is the spatial average of the acoustic energy density~\eqnoref{Eac_alpha} in the cross section $\Omegabar$. Similarly, we use \eqsref{p1_rectangle}{Frad_alpha} to evaluate the axial radiation force,
 \bal \eqlab{Frad_x_rect}
 \frac{F_{\mr{rad},x}^{01}(x,z)}{4\pi a^3 \kbar_0^{01} \avr{\Eacbar^{01}}} \approx   \Big[\frac12 f_1-\Phi\cos^2\big(\kbar_0^{01} z\big) \Big] \sqrt{\Gammabar^{01}}  \pp_\xti  \big|\chi^{01}\big|^2.
 \eal
From \eqsref{Frad_z_rect}{Frad_x_rect}, we calculate the ratio between the maximum axial radiation force and the maximum cross section radiation force for the single standing-wave resonance  by using $\max\big\{\pp_\xti |\chi^{01}(x)|^2\big\} \sim 2\max\big\{ |\chi^{01}(x)|^2\big\}$,
 \bal  \eqlab{Frad_ratio_rect}
 \frac{\max\big\{F_{\mr{rad},x}^{01}\big\}}{\max\big\{F_{\mr{rad},z}^{01}\big\}} &\approx \frac{f_1}{\Phi}  \sqrt{\Gammabar^{01}} = \frac{2}{1+\frac{2 f_0}{3f_1}}  \sqrt{\Gammabar^{01}},
\\ \nn
 &\approx 0.15 \sqrt{\Gammabar^{01}}, \; \text{ for polysterene particles}.
 \eal
 \esub
In the last step we use the scattering coefficients $f_0=0.623$ and $f_1=0.033$ for polystyrene particles~\cite{Settnes2012}. As examples, Ley and Bruus~\cite{Ley2017} studied numerically the pyrex-glass capillary tubes named \qmarks{C1} (used by Hammerstr\"om \etal~\cite{Hammarstrom2012} with inner dimensions $\Ly=2\,\SImm$ and $\Lz=0.2\, \SImm$) and \qmarks{C5} (proposed by the authors with inner dimensions $\Ly=0.5\,\SImm$ and $\Lz=0.2\, \SImm$). In both cases, they found the quality factor for the standing half-wave resonance in the $z$ direction to be $Q=53$, corresponding to the damping coefficient $\Gammabar^{01}=\frac{1}{53}$. Using this value in \eqnoref{Frad_ratio_rect} gives the ratio $\frac{1}{49}$,  where Ley and Bruus found the ratio to be $\frac{0.44\,\SIpN}{22\, \SIpN}=\frac{1}{50}$ for  C1 (see their Fig. 6) and $\frac{0.13\,\SIpN}{7\, \SIpN}=\frac{1}{54}$ for C5 (see their Fig. 9). Hence, even though \eqref{Frad_ratio_rect} is obtained from a hard-wall analysis, it predicts values close to the full simulation where the surrounding glass capillary is included.

Furthermore, for the capillary tube C1 with the experimentally found resonance frequency $f=\frac{1}{2\pi} k_0 \cfl= 3.970$~MHz and transducer length $\Lact=1160 \,\SIum$~\cite{Hammarstrom2012}, we calculate $\Lactti=\Lact k_0 \sqrt{\Gammabar^{01}} =2.6$. From \figref{contours}(a)  and (b) follows the prediction that the acoustic energy density $\Eac^\alpha$ and trapping force $F_{\mr{rad},x}^\alpha$ may be approximately doubled by doubling $\Lact$.

\section{Discussion}\seclab{discussion}

The theory presented in \secref{theory} relies on the main assumption that the 2D resonance modes in an arbitrary cross section can be written as in \eqref{p1_general} for wave numbers $k_0$ very close to the cross-section eigenvalues $\kbar^\alpha_0$. Whereas this generalization is not proven mathematically, it is physically reasonable, and we have validated it in \figref{validate_2D} with a relative deviation of $0.14 \%$. We note that the eigenfunctions $\Rbar^\alpha(y,z)$ in the cross section do not exactly form a complete set, because the eigenvalues of the eigenvalue problem \eqnoref{p1_Eigenvalue} has a small imaginary part. However, in the limit $k_0\deltas\ll 1$ the viscous boundary layer introduces a minute imaginary part to the eigenvalues, and thus
expression \eqnoref{p1_general} is a good approximation. It may be possible to come up with a special cross section where our theory fails, but it does apply to all capillary tube cross sections used in the experiments that are reported in the literature.

We have presented detailed results for the special simplifying, but experimentally relevant, condition that the actuation frequency is near a  resonance characterized by a single mode that does not overlap with other modes, such that the 3D pressure $p_1$ is described by only a single term of the sum~\eqnoref{p1_3d_sol}. We emphasize, however, that this is not a necessary condition, as the general theory allows both for non-resonant actuations and for multiple overlapping modes. In fact, we have done equally successful validations for frequencies away from resonance, where more modes are taken into account.

We have considered the actuation to have a box-shaped axial dependency given by \eqref{psibox} to mimic a piezoelectric transducer confined in the axial direction to a length $\Lact$. In a realistic glass-capillary system, the motion of the wall will be more complicated as found from the numerical simulations by Ley and Bruus~\cite{Ley2017} and the simulations and experiments by Reichert \textit{et al.}~\cite{Reichert2018}. Nevertheless, when calculating the ratio between the axial and cross-sectional radiation force in the end of \secref{halfwave} and using the damping coefficient $\Gammabar^{1}=\frac{1}{53}$ found from the  numerical simulation by Ley~and~Bruus, we almost reproduce their values, namely $\frac{F_{\mr{rad},x}}{F_{\mr{rad},z}} \approx \frac{1}{50}$. This agreement indicates that the predictions from our theory of the axial variations of the pressure remain valid for more complicated wall actuations, and that the important effect from the capillary walls is well described by a change in the damping coefficient $\Gammabar^\alpha$ for the mode. The probable cause for this increased damping factor of the fluid resonance is not dissipation in the capillary tube but instead an axial transport of energy in the solid away from the fluid, as pointed out by Ley and Bruus~\cite{Ley2017}. This can be seen from \eqref{Sac_alpha}, which states that the axial transport $(\SSSac)_x$ of energy is proportional to the speed of sound $\cfl$, and because the speed of sound in the capillary tube is usually larger than in the fluid, energy is efficiently transported away from the trapping region. For example, the longitudinal sound speed in pyrex glass is 3.7 times larger than in water \cite{Corning_Pyrex}.

\section{Conclusion and outlook} \seclab{conclusion}

We have presented a semi-analytical method to calculate the acoustic pressure in a long, straight capillary tube of arbitrary cross section with a localized ultrasound actuation at the walls. Moreover, we have analytically derived the axial dependencies~\eqnoref{time-averaged_quantities} of the time-averaged response and used it to derive an expression for the key aspect in the acoustic trap, namely the axial acoustic radiation force~\eqnoref{Frad_rect} acting on suspended particles. The viscous boundary layer is taken into account through an effective boundary condition~\eqnoref{p1_gov_bc}, which is valid when the width $\deltas$ of the viscous boundary layer is much smaller than both the acoustic wavelength ($k_0 \deltas \ll 1$) and the radius of curvature of the cross section. This condition is usually satisfied in typical experiments.

In \eqref{p1_general}, the acoustic 2D cross-section resonance mode $\pbar_1^\alpha(y,z)$ in an arbitrary cross section was obtained by a generalization of the well-studied case~\eqnoref{p1_rectangle} of a rectangular cross section. The 2D mode $\pbar_1^\alpha(y,z)$ can be found analytically for integrable shapes, such as rectangles, circles and ellipses, and otherwise numerically as shown in \figref{3D_fig}. The theory results in the correct amplitude and phase of $\pbar_1^\alpha(y,z)$ by combining the eigenvalue $\kcbar^\alpha$ and the dimensionless eigenfunction  $\bar{R}^\alpha$ of \eqref{p1_Eigenvalue} with the actuation function $\Ubar_{1\perp}$ of \eqref{p1_gov_U_def} and the inclusion of the viscous boundary layer through the boundary operator $\calD_\perp$ in \eqref{p1_gov_D_def}.

From the 2D pressure modes $\pbar_1^\alpha(y,z)$ in the cross section, known for frequencies near resonance, we derived in \eqref{p1_3d_alpha} the 3D pressure modes $p_1^\alpha(x,y,z)$. The sum over $\alpha$ of these modes, see \eqref{p1_3d_sum}, constitutes the full 3D pressure $p_1(x,y,z)$ including the axial dependency and valid for all frequencies. In \eqref{Lx} we extracted for 3D resonance mode $p_1^\alpha$ the characteristic axial length scale $\Lx^\alpha = \big(\kbar_0^\alpha \sqrt{\Gammabar^\alpha}\big)^{-1}$ , where $\kbar_0^\alpha$ is the 2D resonance wave number and $\Gammabar^\alpha$ is the 2D damping coefficient. Because the axial radiation force $(\FFFrad^\alpha)_x$ and the axial energy flux density $(\SSSac^\alpha)_x$ of resonance mode $\alpha$, depend on the axial gradient of the pressure, we find in \eqsref{Frad_alpha}{Sac_alpha} that these are both proportional to $\sqrt{\Gammabar^\alpha}$. From a purely numerical-modeling point of view, the theoretical method implies a drastic reduction in the memory requirements, because the full 3D system can be obtained from a 2D simulation of the cross-section eigenproblem combined with the analytical expressions for the axial dependencies.

To further study the physics of the acoustic trap, we chose a box-shaped actuation which mimics a piezoelectric transducer attached to the capillary walls in a confined region of length $\Lact$, and which allows for analytic solutions. In \figref{contours} is shown the resulting axial dependencies of the acoustic energy density $\Eac^\alpha$, the cross-sectional acoustic radiation force $(\FFFrad)_{yz}$, the axial acoustic radiation force $F^\alpha_{\mathrm{rad},x}$, and the axial energy flux density $S^\alpha_{\mathrm{rad},x}$. Remarkably, we found an optimal actuation length $\Lact \approx 2\text{-}5 \Lx^\alpha$ that maximizes these quantities. Furthermore, whereas the maximum acoustic energy density $\Eac^\alpha$ is found in the center of the channel, maximum axial radiation force $F^\alpha_{\mathrm{rad},x}$ is located around 17 \% inside the actuation region, and the maximum axial energy flux density $S^\alpha_{\mathrm{rad},x}$ is located at the edge of the actuation region.

We validated numerically our theory in \figsref{validate_d}{validate_L}  for the 3D system shown in \figref{3D_fig} near the resonance $\alpha =1$, by using the box-shaped actuation given in \eqref{psibox}. We found a relative agreement around 1~\% between theory and simulation, even when the box-shaped actuation had smooth steps. This agreement is satisfactory as the theory was developed in the limit of a small boundary-layer width ($k_0\deltas \ll 1$), and $k_0\deltas=0.0024$ in the numerical model.

The presented theory motivates further studies of capillary tubes. One obvious extension of this work is to compute the acoustic streaming, in particular the horizontal in-plane streaming rolls observed by Hammarstr\"om, Laurell, and Nilsson~\cite{Hammarstrom2012} and by Lei, Glynne-Jones, and Hill~\cite{Lei2013}. The streaming can be computed numerically by combining the presented theory with the recently published methods of calculating the time-averaged streaming velocity~\cite{Bach2018, Skov2019}. Another future study, would be to include the elastic walls in the mode analysis. For example, by combining the wall velocity $U_{1\perp}$ obtained from a full 3D numerical simulation with the solution of the 2D eigenvalue problem~\eqnoref{p1_Eigenvalue}, the coupling strength $P_1^\alpha$ for each pressure mode $p_1^\alpha$ can be calculated from \eqref{p1_3d_sol}. In this way the relative importance of each pressure mode in the acoustic trap can be characterized. A last example of further work is to investigate the loss of acoustic energy in the fluid into the solid as briefly discussed in the last paragraph of \secref{discussion}.

We have provided theoretical predictions of the axial variation of the acoustic fields in capillary tubes and pointed out that there is an optimal actuation length leading to a maximum acoustic radiation force, both in the axial and cross-sectional directions. Our analysis provides a theoretical understanding of the complicated 3D characteristics of acoustofluidics in capillary tubes, and in long, straight channels in general. Our resulting expressions can be used to aid in the design of acoustic trapping devices.

\appendix

\section{The Fourier transform and \\the convolution relations} \seclab{Fourier}
We define the Fourier transform $\calF_x$ and the inverse Fourier transform $\calF_k^{-1}$ as
 \bsubalat{DefineFourier_x}{3}
 \eqlab{FourierForward}
 \hat{\phi}(k)&=\calF_x[\phi(x)](k)&&=  \int_{-\infty}^\infty \phi(x)\ee^{-\ii k x}\, \dd x,\\
 \eqlab{FourierBackward}
 \phi(x)&=\calF^{-1}_k[\hat{\phi}(k)](x)&&=\int_{-\infty}^\infty \hat{\phi}(k)\ee^{+\ii k x}\, \frac{\dd k}{2\pi}.
 \esubalat
For this convention, the convolution relations are
 \bsubal{convolution_theorem}
 \eqlab{convolution_theorem_1}
\calF[\phi_1*\phi_2]&=\calF[\phi_1]\calF[\phi_2],
\\
\eqlab{convolution_theorem_2}
 \calF[\phi_1  \phi_2]&=\calF[\phi_1]*\calF[\phi_2],
\esubal
where the in-line asterisk denote the convolution,
 \beq{convolution}
 [f*g](x)=\intinf f(x')g(x-x') \, \dd x'.
 \eeq

\section{Details about the use of the boundary operator $\calD_\perp$ in \eqref{p1_Fourier}}
\seclab{app_calD}
The boundary operator $\calD_\perp$ defined in \eqref{p1_gov_D_def} takes into account the viscous boundary layer. To be able to compare \eqref{p1_Fourier} for $\hat{p}_1$ to \eqref{p1_gov} for $\pbar_1$, we need to express $\calD_\perp$ in terms of $\kc^2-\kx^2$. This is achieved by subtracting and adding $\kx^2$ as $\calD_\perp=\pp_\perp +\frac{\ii}{\ks}\big[(\kc^2-\kx^2)+\kx^2 +\pp_\perp^2\big]$. Close to the poles we have $\kx \approx k_x^\alpha$, and thus  \eqref{kx_alpha} gives $\kc^2-\kx^2\approx (\kcbar^\alpha)^2$, whereby $\calD_\perp \approx \pp_\perp +\frac{\ii}{\ks}\big[(\kcbar^\alpha)^2+(\kx^{\alpha})^2 +\pp_\perp^2\big]$. If we assume $(\kcbar^\alpha)^2\gg (\kx^{\alpha})^2$, we may ignore $(\kx^\alpha)^2$. Later, this assumption is proven correct by noticing that \eqref{kx_size_Gamma} yields $(\kx^{\alpha})^2 \sim \Gammabar^\alpha (\kcbar^\alpha)^2$ with $\Gammabar^\alpha \ll 1$. Consequently, the expression \eqnoref{phat_exp} for the Fourier transform $\phat_1$ at $\kx$-values close to the complex poles $\kx^{\alpha}$ is valid to $\ord{\Gammabar^\alpha}$.

\section{Details in applying the residue theorem} \seclab{Res_details}
To obtain the residues $\mr{Res}\big(\hat p_1(\kx;y,z) \ee^{\ii \kx x},\kx^{\alpha}\big)$ used in \eqref{p1_xyz_Res}, we first rewrite \eqref{phat_int} by inserting $\hat{U}_{1\perp}(\kx;y,z)=\intinf U_{1\perp}(x',y,z) \ee^{-\ii\kx x'} \, \dd x'$,
 \bal \eqlab{phat_int_app}
 &\hat{p}_1(\kx;y,z)\: \ee^{\ii\kx x} \approx
\\ \nn
 &\frac{ -(\kcbar^\alpha)^2 \Rbar^\alpha}{(\kx)^2-(\kx^\alpha)^2}\;
 \frac{\int_{\dOmegabar} \intinf  U_{1\perp}(x',y,z)\ee^{\ii\kx (x-x')} \, \dd x' \Rbar^\alpha\, \dd l} {\int_{\Omegabar} \big(\Rbar^\alpha\big)^2  \, \dd A }.
 \eal
This expression is valid for $\kx$ close to the simple poles given by $\pm \kx^\alpha= \pm \sqrt{\kc^2-(\kcbar^\alpha)^2}$, see \eqref{kx_alpha}. Based on \eqref{phat_int_app}, the integral in \eqref{p1_xyz} is calculated using the residue theorem over a closed contour $\gamma$ in the complex $\kx$-plane chosen as follows: (1) For $x-x'>0$, the integrand  \eqnoref{phat_int_app} vanishes for $\kx \rightarrow \ii \infty$, and we choose the closed contour $\gamma$ to be the counter-clockwise contour consisting of the real $\Re(\kx)$-axis connected to a semicircle of radius $|\kx| \rightarrow \infty$ in the \emph{upper} complex $\kx$-plane. This contour encloses the residues at $\kx=+\kx^\alpha=+\sqrt{\kc^2-(\kcbar^\alpha)^2}$ having positive imaginary part. (2) For $x-x'<0$, the integrand \eqnoref{phat_int_app} vanishes for $\kx \rightarrow -\ii \infty$, and we choose the closed contour $\gamma$ to be the clock-wise contour consisting of the real $\Re(\kx)$-axis connected to a semicircle of radius $|\kx| \rightarrow \infty$ in the \emph{lower} complex $\kx$-plane. This contour encloses the residues at $\kx=-\kx^\alpha=-\sqrt{\kc^2-(\kcbar^\alpha)^2}$ having negative imaginary part. In either case (1) or (2), the residues inside the closed contour $\gamma$ are,
 \bal \eqlab{Res_result}
 &\mr{Res}\big(\hat p_1(\kx;y,z) \ee^{\ii \kx x},\kx^{\alpha}\big) \nn \\
  &=\frac{ -(\kcbar^\alpha)^2 \Rbar^\alpha}{2\kx^\alpha}\; \frac{\int_{\dOmegabar} \intinf  U_{1\perp}(x',y,z)\ee^{\ii\kx^\alpha |x-x'|} \, \dd x' \Rbar^\alpha\, \dd l}{\int_{\Omegabar} \big(\Rbar^\alpha\big)^2  \, \dd A } \nn\\
 &=-\ii  \calGbar^\alpha(\kc)\Rbar^\alpha \frac{\int_{\dOmegabar} \intinf  U_{1\perp}(x',y,z)g(x-x') \, \dd x' \Rbar^\alpha\, \dd l}{\int_{\Omegabar} \big(\Rbar^\alpha\big)^2  \, \dd A },
 \eal
where we inserted $\calGbar^\alpha(\kc)$ from \eqref{p1_general_G} and introduced the normalized function $g^\alpha(x)$ given by \eqref{p1_3d_g}. Finally, using the residues \eqnoref{Res_result} in the sum \eqnoref{p1_xyz_Res}, we obtain \eqref{p1_3d_sol} for the acoustic pressure $p_1$.

%\bibliography{acoustofluidics}
%\bibliographystyle{apsrev4-1-titles}
%\end{document}


\end{document}